# Archiving the Relaxed Consistency Web

Zhiwu Xie[1,2], Herbert Van de Sompel[3], Jinyang Liu[4], Johann van Reenen[2], Ramiro Jordan[2]

[1]Virginia Tech
Blacksburg, VA 24061
zhiwuxie@vt.edu

[2]University of New Mexico
Albuquerque, NM 87131
{zxie,jreenen,rjordan}
@unm.edu

[3]Los Alamos National
Laboratory
Los Alamos, NM 87545
herbertv@lanl.gov

[4]Howard Hughes Medical
Institute
Ashburn, VA 20147
liuj@janelia.hhmi.org

## ABSTRACT

The historical, cultural, and intellectual importance of archiving the web has been widely recognized. Today, all countries with high Internet penetration rate have established high-profile archiving initiatives to crawl and archive the fast-disappearing web content for long-term use. As web technologies evolve, established web archiving techniques face challenges. This paper focuses on the potential impact of the relaxed consistency web design on crawler driven web archiving. Relaxed consistent websites may disseminate, albeit ephemerally, inaccurate and even contradictory information. If captured and preserved in the web archives as historical records, such information will degrade the overall archival quality. To assess the extent of such quality degradation, we build a simplified feed-following application and simulate its operation with synthetic workloads. The results indicate that a non-trivial portion of a relaxed consistency web archive may contain observable inconsistency, and the inconsistency window may extend significantly longer than that observed at the data store. We discuss the nature of such quality degradation and propose a few possible remedies.

## Categories and Subject Descriptors

H3.5 [**Information Storage and Retrieval**]: Online Information Services – *Web-based services.* H3.7 [**Information Storage and Retrieval**]: Digital Libraries – *Collection.* H2.4 [**Database Management**]: Systems – *Distributed databases.*

## General Terms

Design, Experimentation.

## Keywords

Web Archiving, Digital Preservation, Social Network, Consistency.

## 1. INTRODUCTION

The web as we see it today is fast disappearing [33]. Nostalgic sentiment aside, also gone is huge amount of invaluable knowledge. Since 1996 when Internet Archive started to collect and archive web pages, the urgency to preserve the web has received gradual but steady recognition. Today, all countries with high Internet penetration rate have established high-profile archiving initiatives, often involving the national libraries, archives and other government agencies [17][34]. These activities are also coordinated by international collaborations such as the International Internet Preservation Consortium (IIPC). The legal hurdles are being cleared. More than 15 countries have passed and more are actively pursuing legal deposit legislations for web content. With or without strong legal protection, these archives have already accumulated close to 10 PB of web data, providing rich opportunities for data mining and analysis [24]. The potential is unlimited and surprisingly interesting use cases are frequently demonstrated. For example, web archives are routinely being used as evidence in legal battles [18].

Technically, most web archives adopt the crawler driven archiving approach. They deploy archival crawlers such as Heritrix or Nutch to crawl and collect web content. The archives preserve such born digital information in archival formats such as ARC or WARC, and then provide various access channels. Technologies such as Memento [39] are also used to assist easy "time travel" to the past and explore the "collective memory".

Web archiving efforts may be comprehensive, where all the web resources and their representations become the targets for preservation. They also can be selective, where the best effort is made to preserve only the more influential and representative portion to its best possible completeness and accuracy. In either case it is almost impossible to circumvent archiving those global-scale and highly dynamic websites such as major information portals, aggregators, and social networks. After all, they are the focal point of the web, where billions of people spend hours per day not only consuming content from, but also contributing content to. However, technologies used to build these websites can be substantially different from those behind the older, smaller websites, and for this reason they may pose unique challenges for archiving.

In this paper we discuss the potential quality degradation caused by relaxed consistency, which has become a common practice in building large-scale, highly dynamic web applications. More specifically, such degradation refers to the archival deviation from the consistent state of a web application. Exactly what consistency means will be discussed in section 3. Intuitively, a consistent state is the one that all web users should uniformly observe as well as the one that web archives should preserve. If they differ from each other, it would be problematic to claim the web archives as "reliable and unbiased", which are also the conditions set out to admit archived web content as legally binding evidences in trials [18]. This paper specifically addresses the archival quality instead of the archival coverage issue. Due to limited resources, it's not always possible to preserve all the representations of a web resource. But for the ones we manage to archive, we want to make sure they reflect people's collective memory.

It is important to note that the problem we discuss here is different from the archival differences resulting from content negotiation, service localization, or personalization. Those archival differences may only reflect multiple consistent states, each of which is arrived at deterministically under its respective localization or personalization scenario. Such differences will not disappear over time, while the quality problem caused by relaxed consistency is volatile by nature.

To illustrate the problem we conducted an experiment on Sina Weibo, a China based Twitter-like micro-blogging web service. By the end of 2012, Sina Weibo boasted half a billion total registered users and 46 million daily active users. As a study indicates [21], microblogging services like this carry distinctive character of news media, therefore the value for archiving them may be similarly justified as that for archiving CNN or New York Times. All Weibo users, including the potential archival crawlers, naturally expect to be treated equally and receive the same information if they follow the same group of people. To test this assumption, we registered two users, h**** and p******, and let them both follow the top 340 most followed users in Sina Weibo. The number of followers of these Weibo celebrities ranged from about 4 to 46 million.

We opened two different browser windows, one in Google Chrome and another in Firefox, both with empty cache, and then invoked the timeline requests side-by-side for these two users. Contrary to the web user's expectation, the responses are not always the same. Figure 1 shows the screenshots of two cases of discrepancies.

In Figure 1 (a) and (b), the window on the left depicts the partial timeline response for User h**** and the right window for User p*****. We use red rectangles to highlight the messages that have been received by one user but not the other. Except for the missing message, both timelines look exactly the same. Still, from a user's point of view there is no good reason why a message should be missing. According to the timestamps, in case (a), the missing message was created about 16 minutes before the timeline request was invoked, and in case (b) it was created about 5 minutes before. In both cases, both users have also received messages sent by the same user whose message was missing in the timeline requests, although these messages are not depicted in Figure 1. Some of these messages were timestamped before the missing message, and the others were after. This is important because it clearly shows the deviation is not purely caused by the time difference on which these two requests are processed at the server or by the network latency. If it were, the users would not have received messages newer than the missing one. About 10 minutes after the missing messages were detected, we refreshed the timeline windows and in both cases the missing messages appeared.

This example illustrates the archival challenges to be discussed in more detail in the following sections. Namely, there exists a type of web application that reduces consistency; therefore does not always disseminate correct information to all relevant users. Note that an archival crawler does not differentiate itself from any other web user. If we adopt the crawler-driven archiving [26], as the majority of the web archives do, it is now possible for the web archive to take in erroneous information that can be easily refuted.

This phenomenon is relatively new to web archives. At least in theory, the web does not produce such inconsistency during the transmission if the transfer protocol is semantically transparent [14]. When the scale of the application is relatively small and can be easily handled by a single ACID (Atomicity, Consistency, Isolation, Durability) compliant data store, the consistency is most likely also guaranteed at the origin server. The problem only surfaces when the scale of the web application grows beyond the technology that guarantees consistency.

This paper is structured as follows. After providing the related work, we formally define consistency, and then discuss how it is relaxed and what it means to web archiving. In the remainder of the paper we will mainly deal with two questions. First, how much of the relaxed consistency web archive may contain inaccurate information and how to characterize them? Second, given the lack of consistency, what can we do? We approach these questions through a controlled experiment, which is described in section 4. We conclude the paper after giving the experimental results and analysis.

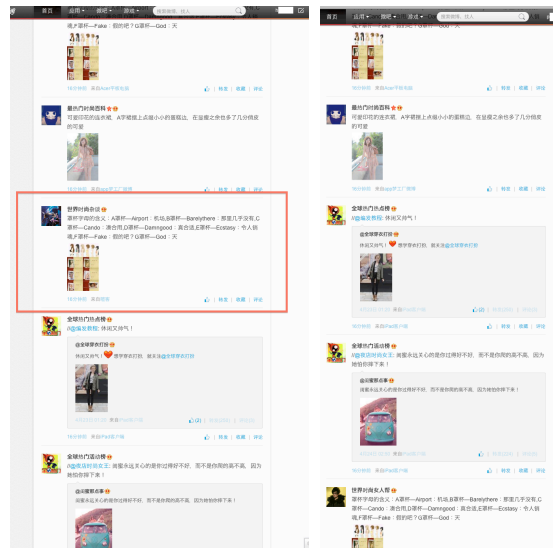

(a)

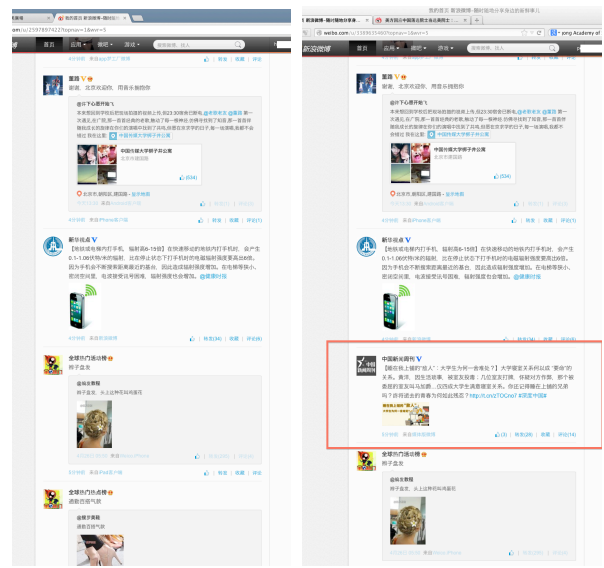

(b)

**Figure 1. Inconsistency observed in Sina Weibo**

## 2. RELATED WORK

This study is related to work done in web archiving, scalability and consistency research in distributed systems and database, and systems research on scaling social networks, particularly the feed following.

Masanès [26] provides a thorough overview on web archiving and describes three major types of web content acquisition methods: the client-side archiving including those based on crawlers, the transactional archiving, and the server-side archiving. Proportional to the archiving practice, the vast majority of the web archiving research deals with crawler-based archiving techniques. A topic of particular interest is how to better detect web page changes and increase crawling efficiency [4][8][28]. Improvements on this front help increasing the archive's coverage of the web. In terms of archiving quality, Denev et al focuses on the sharpness, or the temporal coherence of the archive [12]. The blurring originates from the crawling strategy, which causes the incompatible versions of the interlinked resources being preserved together. Transactional archiving approaches such as SiteStory [40] may be particularly effective to address this issue. This, however, is different from the relaxed consistency discussed in this paper. Here, the quality degradation originates from the inconsistent design of the origin server architecture. Although targeting different problems, the methods and techniques used in the above two distinct bodies of work bear some resemblance to ours and can potentially provide inspiration for future work.

Social network archiving is gaining traction [25], with the focus mainly on Twitter. The Library of Congress started a project to comprehensively archive Twitter [32]. The approach taken was through server-side archiving, where the authoritative server records were to be transferred and preserved. These records included all the tweets but exclude the following network. However this approach is only feasible when the content owner is willing to cooperate. Otherwise crawling remains the next best option. Even for Twitter, the publicly available Garden Hose API only provides a small sample of all the tweets, making the resulting datasets neither comprehensive nor selective. Morstatter et al. point out the limitations of such sampling [27].

Scaling distributed applications has been a hot topic for decades. Traditionally researchers rely on consistency guaranteeing network communication protocols to achieve better performance than two-phase commit. Examples include the multicast total ordering as used in Postgres-R [20], RSI-PC as used in Ganymed [29], the total ordering certifier as used in Tashkent [13], Pub/Sub as used in Ferdinand [15], and the deterministic total preordering [38]. Their performance is in theory upper-bounded by the centralized service implementing these protocols.

The explosive growth of the global-scale web applications, especially the social networks, demands even higher scalability and availability. Relaxing consistency has since gained not only theoretical backing [1][6][16] but also industrial support. Following the seminal papers on Google Bigtable [7] and Amazon Dynamo [11], various relaxed consistency techniques and systems have been developed and widely used. Examples include MongoDB, CouchDB, Cassandra, Riak, Voldermort, and PNUTS [9] etc. Cloud hosted and managed key-value stores like Google App Engine, Amazon SimpleDB, and Amazon DynamoDB further push such technologies to wider market at commodity price. However eventual consistency [41] may not be ideal for all applications. More recent research [2][3][9][22][31] recognizes the need for tighter consistency, e.g., causal consistency. However in the context of archiving, the causality cannot be determined a priori. Any missing message from the historical records may have implications not obvious at the time of archiving.

A number of recent researchers report their results on evaluating inconsistency in relaxed consistency data stores [5][30][42]. The inconsistency windows range from 200ms to 12 seconds. We adopt the observable inconsistency approach from Rahman et al [30], but because we are measuring different things, our results are several orders larger than theirs.

Using relaxed consistency key-value store for social networking functionality is an ongoing effort [23]. Yahoo! PNUTS [35][36] in particular has been used to handle the feed-following problem, which our experiment actively follows. However they have not reported what level of inconsistency has been observed.

## 3. RELAXED CONSISTENCY

In this section we discuss the definition of consistency, why and how large-scale websites relax consistency, and what this means to web archives.

Consistency has different meanings in different contexts. In this paper, we adopt the definition given by the proof [16] of Brewer's Conjecture [6] or the CAP theorem. Under this definition, a consistent system, even built on distributed machines, guarantees an illusion of a total order in which concurrent events can be observed and interpreted as happening on a single machine.

A consistent web service must give all its users a unified view of how things happen on an imaginary single server, no matter if and how they are executed on many distributed machines. Any conflict between the views must be resolvable through the established global order. For example, let us assume a global order of events (i, j, k, l) is established as $i <_t j <_t k <_t l$, where $<_t$ denotes the "happens before" relation. If User A sees $i <_t j <_t k$ and User B sees $j <_t k <_t l$, the difference can be easily interpreted as User A's request is processed at the server before User B. On the other hand, if User A sees $i <_t j <_t l$ and User B sees $j <_t k <_t l$, then the missing event k cannot be easily explained therefore indicates inconsistency.

Maintaining consistency in a large-scale distributed system is very expensive. Moreover, the CAP theorem states that if a network partition occurs, then it is impossible to guarantee both availability and consistency. Even without the network partition, many system designers opt to relax the consistency in order to achieve lower latency [1]. This forms the theoretical basis for relaxing consistency in large-scale web services.

In a relaxed consistency model, the system is allowed to have a period of "inconsistency window" during which a global order cannot be established. For example, in a shared nothing, fully or partially replicated distributed environment, we may declare an event update successful as soon as one of the replicas commits it locally and before this update finishes propagating to the other replicas. By eliminating the consistency locks, the replicas become more independent and can work in a more concurrent manner. But as a result, the system user's view may become rather unpredictable until the update propagates to more replicas than a quorum and the system enters a consistent state. The archival deviation depicted in Figure 1 demonstrates such distinctive characteristics of an ephemeral inconsistency window, which also helps to explain why refreshing can recover the missing messages. Indeed, the Weibo technical team confirmed that their system architecture included various relaxed consistency components [43].

Consistency may be relaxed at multiple subsystems composing the web service. It is impossible to exhaustively enumerate where and how the consistency may be relaxed. The following are a few obvious options. The persistent data store is usually the most apparent candidate to relax consistency since it is often the scalability bottleneck. Most commercial and open-source NoSQL database systems provide such relaxation, often down to the level of eventual consistency [41]. Although some NoSQL and NewSQL database systems boast strict consistency, since they restrict the type of consistent transaction, e.g., to per row [31] or within the same data partition [19], in practice the more complicated queries still need to be broken down into multiple conforming transactions at the application layer. This will compromise the consistency guarantee. Similarly, if an application is backed by a key-value store but the data manipulation can not be easily mapped to a simple key-value READ, WRITE, or SCAN operation, the application layer will still have to introduce more inconsistency. Inconsistency may also be intentionally injected from the application cache layer in order to reduce the server workload. More importantly, the effects of relaxation from multiple subsystems can compound, resulting in an even worse consistency situation than that of any single one.

As the Weibo example shows, the effects of relaxed consistency may seep into web archives and degrade their quality. These effects may include, at least in theory, all identifiable concurrency anomalies. Nevertheless, the relaxed consistency technologies are becoming prevalent in many if not all leading web portals, news aggregators, and social networks. It is often hard to pinpoint which website uses what technology at which level unless disclosed by their technical team.

Given the prominence of these websites, we naturally want to assess the extent of the quality degradation they may cause to the web archives. Another related question concerns the inconsistency window. Prior research has shown that the maximum inconsistency window in many NoSQL data stores is only in the order of seconds [5][30][42], but the Weibo experiment exposes inconsistency delays up to 16 minutes. How do we explain the difference? Is this the norm or exception? We will discuss these issues in the next few sections.

## 4. AN EMPIRICAL STUDY
In this section we describe an empirical study used to assess and characterize the archival quality degradation. We explain the methodology and give detailed descriptions on the experimental settings. The results and analysis are presented in section 5.

### 4.1 Methodology
We propose to gauge the archival inconsistency with a controlled experiment. We choose feed following as the representative web application, and argue that only observable inconsistency needs to be concerned. We further simplify the case so that only two types of inconsistency exist and then force an artificial global order in the experiment in order to significantly cut down the computational complexity for conflict detection.

#### 4.1.1 Controlled experiment
A controlled experiment may be a better or even the only option to seek sensible answers to our questions. This is because existing web archives do not provide sufficient data to expose inconsistency from within, yet it is not quite feasible to conduct large-scale experiments against live web services either.

The frequency in which the existing web archives crawl the web is too low for our purpose. For example, although we know Yahoo! uses relaxed consistency data store PNUTS to power its homepage [37], as late as March 2013 the Internet Archive's Wayback Machine took only 512 snapshots of it for the whole month, averaging about 16 snapshots per day. The crawling frequencies for existing Twitter collections in Archive-It are even lower, averaging about once every few days.

Conducting experiments on live websites can be problematic too. Even if we have access to the backend, it is almost impossible to establish a true global time and global order in a real-world, massively distributed environment. This can be as hard as the original challenge that the relaxed consistency design chooses to circumvent. Moreover, we will not be able collect all the request/response pairs from a live system as well as all the information about the data models, relations, and interconnections. A live system is always changing, making it even harder to detect the inconsistencies. Besides, we will not be able to control the workload and the working conditions of a live web application. Their fluctuations will significantly impact the level of inconsistency.

#### 4.1.2 Feed following
We need to choose an appropriate web application for the controlled experiment. This application should be inherently hard to scale otherwise there will not be much incentive to use relaxed consistency technologies. It should be broadly representative of the real world web applications that handle big data and struggle to meet the needs of large amount of users. Preferably the data model is simple and abstract; it should be easily set up and tested, and allows us to focus on the core inconsistency instead of unrelated issues. The feed following problem seems to be an ideal fit.

Feed following is based on a following network consisting of large numbers of feed consumers and feed producers. Each feed consumer follows a usually large and distinctive group of feed producers, and each producer independently produces event items over time. Now each of the consumers wants to query the *n* most recent event items produced by all the producers this particular consumer follows. Silberstein et al. give a more formal definition of the problem [37].

Feed following is known to be hard to scale [36], yet it forms the foundation of many web portals, aggregators, and social networks. Twitter's timeline application is a typical feed following problem, where each event item is called a tweet. Many other social networking features may also be modeled as variations of feed following, and the "*n* most recent" predicate may also have many other flavors. But the common theme is that each feed following query can be quite personalized and distinctive from the others.

Feed following is the target of many relaxed consistency researches. A naïve relaxed consistency solution is to build a materialized view for each consumer reflecting its changing timeline [23]. When a producer sends a new tweet, the system will preemptively update all timeline records for consumers following this particular producer. When a consumer requests its timeline, the system will directly respond with the established materialized view with no further database query. The query latency will be very low although it takes more time to process a new tweet. Since updating large number of records in one atomic transaction is expensive, the relaxed consistency approach chooses to abandon the atomicity requirement and allows the updates to be conducted asynchronously. For example, as soon as any of the timeline updates is successful, the system could declare the tweet event successful and move on to handle the other requests. The application layer will keep updating the other timeline views and the key-value store will keep replicating these updates across all the database replicas. But if any of these consumers now issues a timeline request and the request lands on a replica that has not been updated, a potential inconsistency is produced and propagated to the end user. In this circumstance, the source of inconsistency is not limited to the data store. Since the application layer breaks down a supposedly atomic transaction into potentially large number of key-value operations, the resulting inconsistency may be much higher than that caused by the inconsistent key-value store. Whether the inconsistency can be observed depends on the other users. We will discuss this topic in more details in the following sections.

### 4.1.3 Simplifications

To reduce the complexity of the analysis, we introduce two simplifications. First, we keep the established social network unchanged during the experiment. This eliminates all the conflicts caused by mismatches between the timelines and the changing following network. It also drastically simplifies our data model design and allows us to replicate the full following network to the front-end servers. Second, we do not allow tweet deletions and retweets. Now all updates in the system are new tweet events and the only possible cause for inconsistency is missing tweets.

### 4.1.4 Archival Quality and Inconsistency

Since the archival crawler does not differentiate itself from the other system users, we can treat the archived content as an unbiased sample from all the user responses, therefore the inconsistency rate of all the user responses may be used as an indicator of the archival quality.

Rahman et al [30] argue that inconsistency measurements should take the client-centric view and avoid the observer's effects. Following the same rationale, in this experiment we only focus on the observable inconsistencies, which are the user responses that conflict with each other. The purpose is to ensure that as a minimum the web archive does not contradict with what people can see. This distinction is important, because not every possible web system state has been exposed to human consumption. For archives that collect people's collective history and memory, we can safely regard those system states as nonexistent if they have never been seen by anyone. If they do not exist, there is no contradiction and no inconsistency.

Due to our simplifications, the types of observable inconsistencies may be explicitly identified. As shown in Figure 2, assume both User 2 and 3 follow User 1, and User 1 sends new tweets in the following sequence: $A <_t B <_t C <_t D <_t E$. Now both User 2 and User 3 issue timeline requests. User 2 receives a timeline response that consists of A, C and D. The response is timestamped at $t_i$. User 3 also receives a timeline response consisting of A, B, and C, but is timestamped at a later time $t_{i+1}$. We identify the following two types of conflicts:

- User 2 does not see B in between A and C. If there exists any other timeline, e.g., timeline(User 3), that contains B, then timeline(User 2) is considered inconsistent.

- User 3 does not see tweet D at time $t_{i+1}$. If we can find any other timeline, e.g., timeline(User 2), that was timestamped earlier than $t_{i+1}$ and it contains D, then timeline(User 3) will be identified as inconsistent. We cannot blame the disappearing tweet on the network latency, because consumer 2 sees it even before the frontend server responds to consumer 3.

Note that neither staleness nor network latency causes conflict in our experiment.

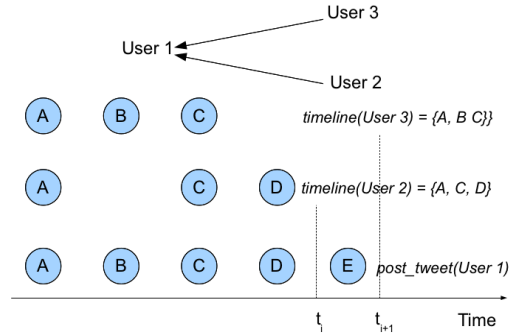

**Figure 2. Detect Inconsistency**

### 4.1.5 Establishing normal working condition

As Rahman et al [30] point out, in a relaxed consistency system, a higher workload will stress out the system and exacerbate the inconsistency. Some prior work detects inconsistency on individual key-value pair by applying extreme workloads [5]. In the context of web archiving, what we want to know is not the system's inconsistency limits under the worst-case scenario. Instead, we are more interested in the inconsistency level observed under the normal operating conditions. More likely these are the conditions under which the archival crawlers gather web content. Of course, there are many interpretations on what constitute a normal operating condition. In this paper we use a benchmarking tool to establish such working conditions.

### 4.1.6 Detecting inconsistency

Even after substantial simplifications, detecting inconsistency can still be an intractable problem. We would like to avoid having to crosscheck inconsistency among millions or more timeline responses. Realizing that the observable inconsistent timeline as a whole is a subset of all inconsistent timelines, we devise a method to first shortlist the possible inconsistent timelines, and then only compare them with the other timeline responses for inconsistency. This, however, requires establishing a global order.

Although establishing the global order is difficult in a real world system, it is indeed possible in an experimental setting. We take lessons from Thomson and Abadi [38] and force all new tweets to be submitted to a single frontend machine. This is the machine that assigns timestamps from its local system clock that forms the global order. After the timestamp is assigned, the request is then sent to the backend data store. Even if the request is unsuccessful and needs to be redone, the established timestamp does not change in the process. We put the user_id and the timestamp directly into the message body in JSON format, as shown in the following:

```
{"producer_id": "1353955", "t": "2013-01-31T04:00:32.256647"}
```

This allows us to easily compare timelines for inconsistency detection.

## 4.2 Experiment Configuration

We now conduct the feed following experiment. We first establish the following network and the workload based on a Yahoo! PNUTS based feed following experiment. We then build a feed following system and run it on Amazon EC2. We choose Amazon DynamoDB as the backend data store and choose per-key strong consistency as our level of relaxed consistency. After running the experiments, all logged data are transferred to and processed on another cloud application built for the purpose of detecting conflict.

As explained before, we assume the conflict rates are the same for the archived contents and the responses received by all the users. We therefore skip the crawling and the archiving steps without affecting the validity of this experiment.

### 4.2.1 Following Network and Workload

We derive both the following network and the workload from a Yahoo! PNUT based feed following experiment [37]. We assume these can be used to represent a typical web application under normal working condition. The Yahoo! experiment built the following network by crawling Twitter and then derived the workload from the Yahoo! Social Updates platform. They concluded that both the following network and the workload followed Zipfian distribution. We therefore adopt the same Zipf parameters but use Yahoo! Cloud Serving Benchmark [10] to generate a synthetic following network as well as the workload, as shown in Table 1. The slight differences between the average values indicate the different sources of data: Yahoo!'s data were collected from real-world applications but ours are generated from a benchmarking tool. In our following network of about 200,000 users, the most popular producer has 5452 followers, and the nosiest consumer follows 335 producers, although the average numbers are only 13.38 and 4.63, respectively.

**Table 1. Comparing workload parameters with the Yahoo! PNUTS based feed following experiment [37]**

|  |  | PNUTS | This |
|---|---|---|---|
| Number of producers |  | 67,921 | 67882 |
| Number of consumers |  | 200,000 | 196,283 |
| Consumers per producer | Average | 15.0 | 13.38 |
|  | Zipf parameter | 0.39 | 0.39 |
| Producers per consumer | Average | 5.1 | 4.63 |
|  | Zipf parameter | 0.62 | 0.62 |
| Per-producer rate | Average | 1/hour | 1/hour |
|  | Zipf parameter | 0.57 | 0.57 |
| Per-consumer rate | Average | 5.8/hour | 5.8/hour |
|  | Zipf parameter | 0.62 | 0.62 |

### 4.2.2 Consistency level

A crucial piece of this experiment is the relaxed consistency data store that demonstrates properties of relaxed consistency. In the early stage of this study we decided to run the experiments on a computing cloud. DynamoDB, Amazon's newest generation of managed key-value store, was launched in early 2012 to coincide with this experiment.

Since DynamoDB is managed, we do not have to tweak the configuration in order to get the best performance or risk skewing the results. Built on novel ideas from many other key-value stores and researches, DynamoDB actually offers consistency options tighter than pure eventual consistency. After thorough consideration, we decided to take advantage of the per-key strong consistency feature, namely conditional WRITE. A conditional WRITE checks to make sure the value to be overwritten is indeed the one it is supposed to be. Using this feature, new tweets are written into individual follower's timeline in a strictly ordered manner across all the replicas and no new tweet can be permanently overwritten due to inconsistency. We consider this a base requirement for any similar web application, although it costs twice as much as an eventual consistency WRITE. A failed WRITE also causes our implementation to retry until it succeeds. This change intentionally avoids many detectable inconsistencies that could have happened in a pure eventual consistency implementation. The overall consistency property, however, remains eventual consistency. This is because each new tweet still involves many timeline WRITEs and as a whole they are not completed in a single, atomic transaction.

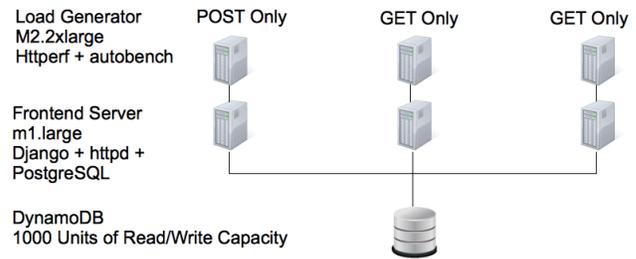

**Figure 3. Feed Following Experiment Configurations**

### 4.2.3 Server configuration

The experiment is conducted on the Amazon clouds. The server configuration is shown in Figure 3. To ensure the results are not skewed by limited computing resources, we provisioned sufficient machine and database capacities, much higher than they normally require. We have not observed any overload during the experiments.

We provision three pairs of servers. Each pair consists of one httperf server that emits the workload and one frontend server that runs a Django application that implements the feed following model. One pair is only used to post new tweets. The frontend machine in this pair timestamps, serializes, and logs all the WRITEs by its local system clock. The other two pairs serve the timeline query workload. No response is cached during the experiments.

We run the experiment for a little longer than two hours. Both the consumed READ and WRITE capacities on DynamoDB reached the level of about 400 READ/WRITE per second. All the query responses are logged, totaling about 2 Gigabytes.

### 4.2.4 Data processing configuration

We anticipate more conflict will appear in the later stages of the experiment, because in the initial stage most timelines are still empty. After the initial warm-up stage these timelines start to get filled and repeatedly updated, that is when inconsistencies become more visible. We therefore pick the data from the latter half of the experiment, totaling 1.2 million timelines.

For each of these timeline results, we must first calculate the consistent timeline from the total order and the following network. This would require repeated database queries against the following network table. For each of the missing tweets discovered, we must then ask the question: which consumer has ever received this missing tweet? The second type of conflict detection also requires comparing the timestamps. This requires large amount of the processing power and database throughput, which forces us to build another data processing cloud infrastructure for this purpose. We first provision three X-large EC2 instances; each handles a portion of the timeline response result. These machines parse the results, and then load the timeline data into another DynamoDB table, which takes several hours. We repeat the data loading for three times to ensure all tweets in the timelines are in DynamoDB. We then provision 100 small EC2 instances; each runs its own local PostgreSQL database containing the following network relations, and a Python script to query the DynamoDB through the boto library. This portion of data processing takes about two hours.

## 5. RESULTS AND ANALYSIS

In this section we report the experimental results and provide some basic interpretations. As much as 6.27% of the responses contain observable conflicts, and on average they are observed 823 seconds after the missing messages are created. This delay is much larger than the inconsistency window measured at the data store, indicating the majority of the inconsistency is most likely created at the application layer.

We further analyze the data in an attempt to correlate the inconsistency with the properties of the consumers and the producers. The results show that at least in our implementation, inconsistency only positively correlates with the producer's popularity, implying that the Internet celebrities' tweets are more likely to be subject to inaccurate web archiving.

### 5.1 Level of Inconsistency and Time Gap

Out of the 1.2 million timeline responses we analyzed, a total of 75,181 responses or 6.27% contain observable conflicts. This is a non-trivial percentage, indicating the problem discussed in this paper cannot be easily dismissed as marginal.

We now take a closer look at their temporal properties. Suppose at the consistency detection stage we discover $m$ missing tweets that belong to a consistent timeline timestamped at T. Among these $m$ tweets we can further identify observable conflicts on $n$ tweets, say $M_0, M_1, \ldots M_{n-1}$, and we have m >= n. Let $T_i$ be the timestamp of the ith tweet, for $0 <= i < n$. We can be certain that $T_i <_t T$ for all i, otherwise the timestamps are mistaken that let people see things in the future. We define the Inconsistency Time Gap

$$G = \max(T - T_i)$$

Note G is not the inconsistency window but is bounded by it.

We now calculate G for each of the 75181 observed conflicts. To our surprise, all except 2 have G greater than one second. Moreover, the average G is as high as 823 seconds, roughly in line with the 16 minutes inconsistency gap detected in the Weibo experiment. An hour-long inconsistency gap is not rare.

The distribution of the G value is depicted in Figure 4. In the figure, axis x denotes the distribution range and axis y denotes the number of G value that falls in the range. For example, 9731 G values fall in the range from 0 to 99 seconds, 7693 G values fall in the range from 100 to 199 seconds. In comparison, only 79 G values fall in the range from 4000 to 4099 seconds, and 7 in the range from 6000 to 6099 seconds.

As mentioned in section 2, existing work put the measured inconsistency window in the order of seconds, yet the time gap in this experiment is several orders larger than that. This certainly begs explanation. We believe the reason lies in the difference on what we measure. We measure overall inconsistency at the application layer for a rather hard-to-scale application, but their experiments are focused on the data store, mostly on single key-value READ/WRITE. Since the feed following cannot be easily mapped to any key-value store's data manipulation primitive, enormous inconsistency is introduced at the application layer and compounded to the data store inconsistency. We anticipate many real world hard-to-scale web applications will be in the similar situation as ours, as demonstrated by the Weibo experiment.

This distribution leads us to believe that inconsistency is a tangible problem for archiving relaxed consistency web content. Moreover, it clearly illustrates that the inconsistency level indeed decreases significantly with time. The promise of eventual consistency has been kept, although the wait can be fairly long.

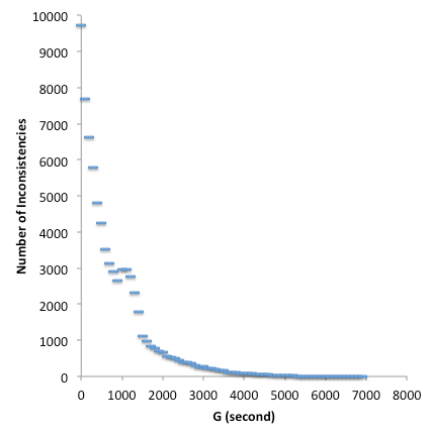

Figure 4. Inconsistency Time Gap

### 5.2 Inconsistency and Producers

Does inconsistency vary with the producer properties, such as, how popular and how active they are? Figure 5 shows the experimental results.

It comes as no surprise that in this particular implementation a strong correlation can be found between the producer popularity and the inconsistency she causes, as clearly shown in Figure 5(a). To better illustrate the correlation we plot the results in logarithm scale, with the x-axis denotes the logarithm of the number of followers a producer has. Since all conflicts can be attributed to missing tweets, we can easily trace them back to the offending producers then the total number of inconsistency the producers cause.

The strong correlation may be explained as follows. In our implementation, when a producer, e.g. user #1353955 who has 234 followers, sends a new tweet, we must asynchronously insert the following line into all 234 key-value pairs, each containing one of its followers' timeline.

    {"producer_id": "1353955", "t": "2013-01-31T04:00:32.256647"}

The more followers, the more views need to be updated and maintained, and the longer it takes to reach a consistent state, therefore the higher possibility of inconsistency.

Such correlation exacerbates the archival problem, because the Internet celebrities' tweets carry more weight as news media and tend to have higher preservation value.

On the other hand, as Figure 5(b) illustrates, there is little or no correlation between inconsistency and how active a producer is. This implies that directing the archival crawlers away from the active producers may have little effect on archiving quality.

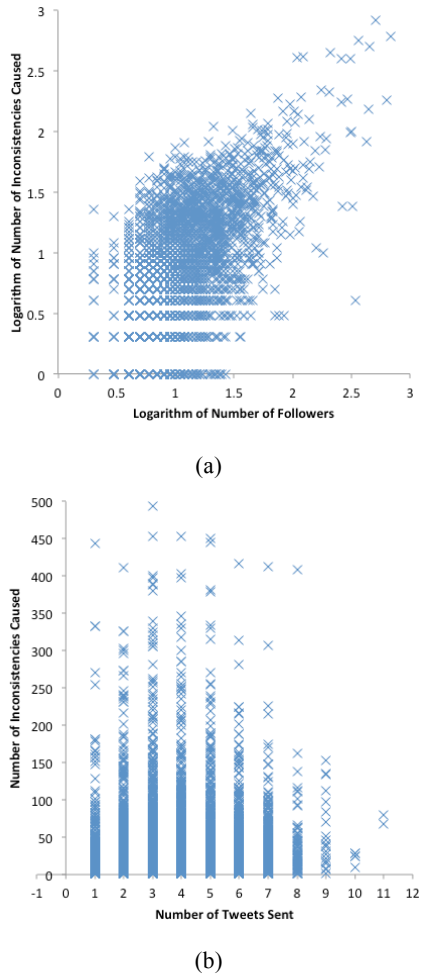

(a)

(b)

**Figure 5. Inconsistency and Producer**

### 5.3 Inconsistency and Consumers

We now attempt to establish correlations between the inconsistency and the consumer's behavior. We may think of the feed consumer in this case as a representative information consumer in any web environment, including the archiving crawlers. Can they change their behavior to circumvent inconsistencies?

We focus on two specific properties of the consumer: the number of producers she follows and the frequency she makes timeline queries. Intuitively we anticipate more active information seekers may encounter higher percentage of inconsistencies, but the experiment results do not seem to support this conjecture, at least under our experiment settings and implementation. Figure 6 gives the results of these two relations. In Figure 6(a) the x-axis denotes the number of producers a consumer follows. The y-axis denotes the number of inconsistencies this consumer encounters. The figure shows no obvious correlation between the two. The same is true for the other factor, namely the number of timeline requests a consumer made, which is depicted in Figure 6(b).

Such results may be counterintuitive because in a real world social network there may exist some correlations between the activeness of the producers and the consumers. However, the following network and workloads used in this experiment were not deduced from real world following applications as the Yahoo experiment. Instead, they were synthesized using the same Zipf parameters. The activeness correlation is therefore lost. Nevertheless, such loss of reality turns out to be advantageous for our purpose, because it allows us to separate the spillover effects of the producer activeness from the true causality.

Overall, the results lead us to lean more towards the belief that simply adjusting the crawling coverage or frequency may not necessarily improve the archival quality.

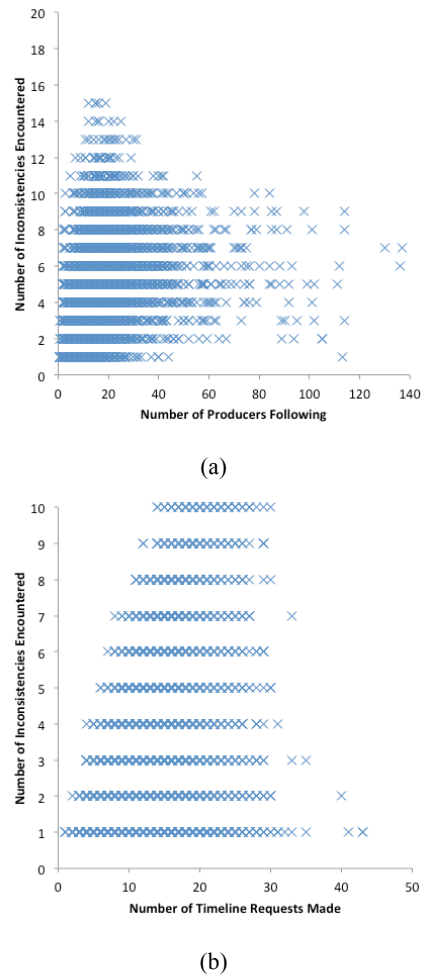

(a)

(b)

**Figure 6. Inconsistency and Consumer**

### 6. CONCLUSIONS AND DISCUSSION

In this paper we explore the archival quality degradation associated with crawling the relaxed consistency web applications. The archived content is error-prone largely by design. To gauge the size of the problem and gain insight into possible solutions, we conducted a controlled experiment and gathered data for analysis. We can draw the following conclusions from the experiment:

- This study confirms the evident and non-trivial presence of archival quality degradation.
- The inconsistency window may be significantly larger than previously reported results. This observation is also supported by cursory tests on an existing relaxed consistency web service.
- The inconsistency level decreases significantly with time. The promise of eventual consistency has been withheld, therefore may be leveraged to offset the quality degradation.
- At least under certain circumstances, it may be harder to capture an accurate snapshot of the more popular web resource.

However, we would like to urge caution on extrapolating the experimental results much further from their contexts. For example, it would be premature to conclude that consumer behavior has absolutely no effect on archival quality. In fact, the Yahoo! PNUTS based feed following implementation [37] does not build a materialized view for every follower. Instead, a cost function is established for each consumer/producer pair to decide if a materialized view is necessary. In this case the consumer behavior most likely will affect the resulting inconsistency.

Now that we can confirm the presence of the archival inconsistency, what can we do? In the following we propose a few possible ways to approach future work.

A proactive approach is to set up multiple archival crawlers to conduct redundant crawling. Like our Weibo experiment, we may set off multiple crawlers to crawl the same resource at the same time; we then compare the results and filter out the possibly inconsistent responses. As pointed out by Bailis et al. [3], inconsistency by nature is instable and probabilistically bounded. Depending on the number of replicas used in the data store, the replication strategy, and how the data models are handled in the key-value design, we may be able to determine how probable it is for a crawler to receive consistent responses within a certain period of time. We may then decide how many crawlers we'll need to increase the probability to an acceptable level. Even without the exact knowledge of the probability, using multiple crawlers should still improve the archival quality, since it's highly improbable for the inconsistent responses to be exactly the same as each other. This approach, however, has its limitations. For example, in general content owners do not allow highly active crawlers, especially when they intensively crawl the web resources in narrow time spans. The server may also choose to flatten such peak workload by delaying some of the processing, which defeats the purpose of redundant crawling.

A compensatory approach is to locate the possible inconsistent copies, run consistency check, then label, remove, or modify them. For example, we may heuristically figure out the inconsistency window, possibly by doing analyses like ours and then deduce it from the G value. With this knowledge in hand, we can be relatively certain about which archived copies are less likely to be inaccurate. These are the copies we can preserve for long term. For the rest, we may discard the fresher part of the responses, verify or recover them from other collected content, or actively re-crawl the resource in order to correct the previously archived one. As we mentioned in section 2, such re-crawling has been widely used to both broaden the archival coverage and correct the temporal incoherence. Alternatively, we may simply label them as possibly inaccurate, assign credibility scores based on inside knowledge about the systems, or hold them as valid unless proven otherwise. This latter approach also applies to the existing archived content where crosschecking or re-crawling is no longer feasible.

The effectiveness of the above approaches still needs further experimental validation. As the last resort, we can always fall back to server-side archiving on important web resources.

To conclude, this study explores a potentially substantial yet previously overlooked web archiving problem. The insights gained may help the web archives to adjust the strategy and improve the archival quality.

# 7. ACKNOWLEDGMENTS
The authors wish to thank Dr Martin Klein for his feedback on a draft version.